\documentclass[aps,amsmath,amssymb,onecolumn,notitlepage]{revtex4-2}
\usepackage{graphicx}
\usepackage{bm}

\usepackage{makecell}
\usepackage{float}
\usepackage{color}   
\usepackage{siunitx}
\usepackage{hyperref}

\definecolor{linkcolor}{rgb}{0,0.5,0.73}
\definecolor{red}{rgb}{0.6078,0.1098,0.1922}
\definecolor{orange}{cmyk}{0,0.386,0.730,0.055}
\definecolor{green}{rgb}{0.4667,0.7294,0.6}

\begin{document}
\title{Universal scaling of brine migration in ice}
\date{\today}
\author{Natanaël Braos--d'Urso, Nicolas Bain, Bruno Issenmann and Sylvain Deville}
\affiliation{Université Claude Bernard Lyon 1, CNRS, Institut Lumière Matière, UMR5306, F-69100, Villeurbanne, France}

\begin{abstract}
Sea ice hosts mobile brine inclusions whose migration influences salt redistribution, microstructure and transport within the ice. Although their motion under temperature gradient is commonly described as a diffusion-limited process, experimental evidence has remained rare and insufficient because migration is slow and controlled inclusion seeding is challenging. Here we introduce an experimental method to trap and monitor hundreds of inclusions in ice under controlled temperature gradients. Using four solutes, three imposed gradients and two order of magnitude in initial concentration, we quantify migration dynamics over a broad range of conditions. We find that, except close to the ice front, inclusion migration is consistent with a regime limited by solute diffusion. After applying a universal scaling, both our data and previous measurements collapse onto a master curve, providing a general framework for describing solute transport in ice.
\end{abstract}

\maketitle

When you contemplate sea ice, it seems quiescent. But this stillness is an illusion. Sea ice is filled with a multitude of brine inclusions and channels in motion which play a key role in the polar ecosystems and in climate. Brine inclusions form during the freezing of seawater, when growing ice rejects salt ions and traps them into liquid pockets~\cite{galley2015}. Once trapped, they are dynamic and migrate through the ice under the influence of thermal gradients and gravity~\cite{hoekstra1965migration,worster1997natural}. With time, inclusions can merge to form brine channels, significantly altering the ice microstructure and permeability. Brine inclusions and channels then play a crucial role in regulating the salinity, mechanical strength, and thermal properties of sea ice~\cite{cole2001microstructure,schwerdtfeger1963thermal}, while also facilitating the exchange of gases, nutrients, and pollutants between the ocean and the atmosphere~\cite{jones2023}. Overall, their distribution and dynamics directly impact habitat availability for ice-associated organisms and the overall stability of polar ecosystems~\cite{lizotte2003microbiology}, as well as sea ice albedo~\cite{carns2015salt}, ice melt rates, and ocean-atmosphere interactions. Understanding these processes is therefore vital for the protection of polar ecosystems and accurate climate modeling.\\

\noindent
Yet, we only have a coarse understanding of brine inclusion migration under temperature gradients in ice. Although we model it as an out-of-equilibrium process called temperature gradient zone melting~\cite{hoekstra1965migration}, where the salt in the inclusion diffuses from the cooler freezing side to the warmer thawing side, experimental confirmations are surprisingly scarce.

They have in fact been limited by three main challenges. First, typical migration speeds of a few $\unit{\nano\meter}/\unit{\second}$ \cite{hoekstra1965migration, light2009hydrohalite} require long and technically challenging experiments. Second, identifying and tracking brine inclusions in natural sea-ice is extremely difficult~\cite{zhao2024ion}. At last, previous studies could only observe naturally present \cite{light2009hydrohalite} or manually injected \cite{hoekstra1965migration} brine inclusions, therefore with limited statistics. Altogether, these restrictions prevent a solid comprehension of brine inclusions migration in ice, even in the simplest cases.

Here, we overcome previous limitations and investigate the migration dynamics of hundreds of brine inclusions in ice. Leveraging a novel method to systematically trap numerous brine inclusions, we optically track their migration under three temperature gradients, for four different salts. We show that, except in a narrow region near the ice front, the migration speed of brine inclusions is limited by the diffusion of salt within the inclusion, with all present and past results collapsing onto a master curve.

\begin{figure}[htbp]
    \centering
    \includegraphics[scale=1]{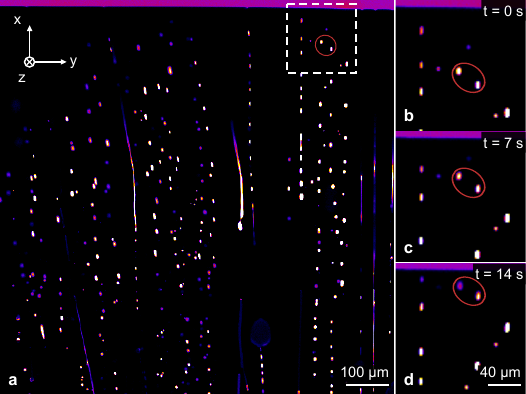}
    \caption{\textbf{Brine inclusions trapped in ice}.
    (A) Full field of view of the imaged sample, in which dozens of inclusions are trapped. Ice appears black, while liquid KCl brine inclusions appear bright due to the dye fluorescence. The water/ice interface is visible at the top of the image. (B) to (D) Enlarged views of the boxed region (A) at different time steps, illustrating the spatial migration of two liquid inclusions over time.}
    \label{fig:method}
\end{figure}

To reproducibly seed a block of ice with large numbers of brine inclusions, we take advantage of an ice-front destabilisation process (Fig.~\ref{fig:method}A).
More specifically, we fill a Hele-Shaw cell with a saline solution of salt at concentration $c_0$ and a small amount of fluorescent dye for visualisation  (\hyperref[sec:exp_setup]{\textcolor{linkcolor}{Methods 1.1}}). We place it in a controlled temperature gradient to trigger ice nucleation, and pull the sample at constant speed to destabilise the ice front into a cellular regime~\cite{dedovets2018} (\hyperref[sec:exp_setup]{\textcolor{linkcolor}{Methods 1.2}}). Due to their insolubility in ice~\cite{petrenko2002}, salts (and the dye) concentrate at the ice front and in liquid films between adjacent ice cells. These films eventually thin out and destabilize into arrays of brine inclusions~\cite{dedovets2018}. The inclusions obtained  with this protocol measure from $70\,\unit{\micro\meter}$ close to the ice front to $10\,\unit{\micro\meter}$ far from it, with typical densities of 50 detected inclusions per mm$^2$ (Fig.~\ref{fig:method}A).\\

\noindent
Once the inclusions are formed, we stop pulling on the sample and maintain it in a constant temperature gradient $G_0$, rapidly reaching a stationary state where the front is immobile.  As the inclusions migrate toward the warmer side (Fig.~\ref{fig:method}B,C,D), we use time-resolved fluorescent confocal microscopy and detection-tracking techniques to measure individual migration velocities (\hyperref[sec:exp_setup]{\textcolor{linkcolor}{Methods 1.3 and 1.4}}). A typical experiment lasts between 15 to 20~minutes and results in hundreds of trajectories (Supplementary Movie).  With this approach, we reproducibly form hundreds of brine inclusions in ice, for different salts at different initial concentrations, and track their dynamics under different imposed temperature gradients.\\

\begin{figure}[htbp]
    \centering
    \includegraphics[scale=1]{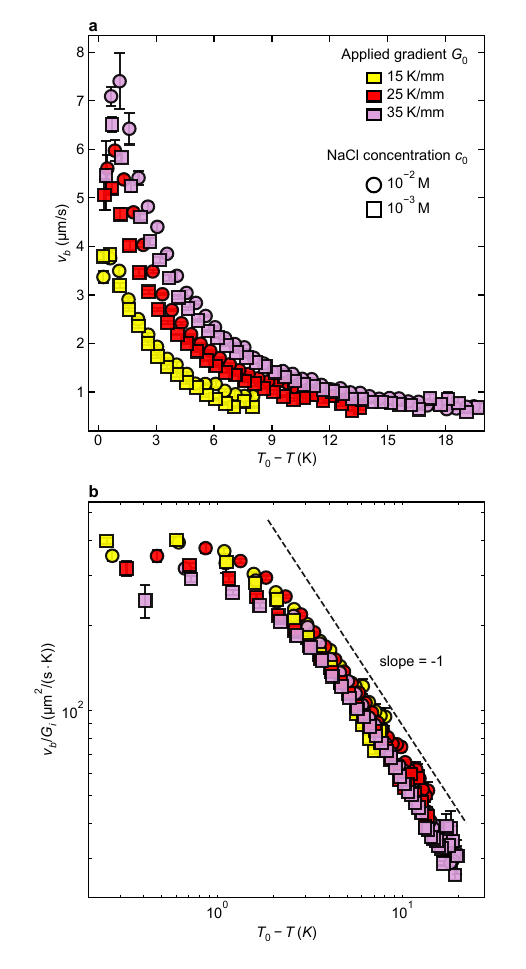}
    \caption{\textbf{Migration velocities for different initial NaCl concentrations and temperature gradients}.
    (A) Migration velocity as a function of local temperature. 
    (B) Migration velocity normalised by the imposed temperature gradient $G_i$ as a function of the local temperature. 
    The yellow, red, and purple colors correspond to the different imposed temperature gradients $G_0 = 15, 25$, and $35 \unit{\kelvin}/\unit{\mm}$ respectively.
    The circles and squares correspond to inclusions prepared from NaCl solutions of initial salt concentration $c_0 = 10^{-2}$ and $10^{-3}$~M, respectively. 
    The position of the data points correspond to the migration velocity averaged over all the trajectories over regularly spaced spatial bins. 
    The size of the error bars equate the standard error ${\sigma}/{\sqrt{N}}$ (\hyperref[sec:vel]{\textcolor{linkcolor}{Methods 3}}).}
    \label{fig:figure_NaCl}
\end{figure}

\noindent
We first focus on NaCl, the most abundant salt in sea water~\cite{lyman1940composition,Millero2008}. Using the protocol described above, we create brine inclusions from NaCl solutions of initial concentrations $c_0 = 10^{-2}$ and $10^{-3}\unit{M}$, and observe their migration under three temperature gradients $G_0 = 15, 25$ and $35\,\unit{\kelvin}/\unit{\milli\meter}$. Assuming ice temperature varies linearly with the distance from the ice front, we quantify the brine migration velocity $v_b$ as a function of local temperature $T_0 - T$ (Fig.~\ref{fig:figure_NaCl}A, \hyperref[sec:temp]{\textcolor{linkcolor}{Methods 2.1}}), with $T_0 = \qty{273.15}{\kelvin}$ the freezing temperature of pure water. We emphasize that each data point does not correspond to the velocity of a single inclusion, but to the velocity of dozens of trajectories averaged over a spatial bin (\hyperref[sec:vel]{\textcolor{linkcolor}{Methods 3}}).
\\

\noindent
We recall that the brine inclusions migrate while being  subjected to a constant and uniform temperature gradient. As expected, the stronger the gradient the larger the migration velocity. Two behaviors are however more surprising. First, in contrast with particles in a fluid under a constant external field, migration velocities here keep increasing when approaching the ice front. Second, although we expect migration velocities to be controlled by the amount of salt, they are nearly insensitive to the salt concentration $c_0$ in the initial solution.
\\

\noindent
To explain these observations, we delve into the mechanisms governing the migration of brine inclusions in ice~\cite{tiller1963migration,pawar2022analysis}. When we generate a brine inclusion, we trap a fixed amount of salt within a brine pocket. In the absence of a temperature gradient $G_0 = 0$, the inclusion grows or shrinks through phase changes at its boundaries, until the salt concentration reaches an equilibrium value $c^*(T)$, where $T<T_0$ is the outer temperature. Under a temperature gradient $G_0 \neq 0$, the system is driven out of equilibrium and the inclusion moves towards the warmer temperatures (Fig.~\ref{fig:method}). This motion is only possible if sustained by a flux of salt, continuously carrying the salt expelled from the freezing side to the thawing side. Brine migration in ice is therefore controlled by heat and salt transport.\\

\noindent
At this stage, three comments are in order. First, thermal diffusivity of liquid water $D_T$ is much larger than the diffusion coefficient $D$ of NaCl, $D_T / D \sim 10^{4}$~\cite{james1968thermal,vitagliano1956diffusion}.
The temperature inside the inclusion is thus a fast variable and takes the form $T = T_b + G_b\tilde x$, where $\tilde x = x - x_b$ is the position relative to the inclusion center $x_b$, $G_b$ the temperature gradient inside the brine inclusion, and $T_b$ the temperature at the center of the inclusion (\hyperref[sec:temp]{\textcolor{linkcolor}{Methods 2.2}}). Second, phase change dynamics are faster than our experimental time scales (\hyperref[sec:phasechange]{\textcolor{linkcolor}{Methods 4}}). The salt concentration along the inclusion boundary is therefore equal to the equilibrium concentration $c^*(T)$. Third, the diffusion time across the inclusion cross section is small compared to the migration time scale (\hyperref[sec:radial]{\textcolor{linkcolor}{Methods 5}}). The salt concentration is then equal to the equilibrium concentration $c^*(T)$ everywhere inside the inclusion.
\\

\noindent
To estimate the flux of salt within the inclusion, we consider the conservation law $\partial_t c + \bm\nabla\cdot(\mathbf j_c) = 0$, where $\mathbf j_c = c\mathbf v - D\bm\nabla c$ is the concentration flux, composed of an advective and a diffusive part. Assuming a linear phase transition curve $c^*(T)$, a constant diffusion coefficient $D$, and a steady inclusion length, a balance between the concentration flux inside the inclusion and the flux of salt due to the migration leads to
\begin{equation}
    \frac{v_b}{G_i} = \mathcal A^2_\lambda \frac{D}{T_0 - T},
    \label{eq:migration_speed}
\end{equation}
where $G_i$ is the temperature gradient within the ice, and $\mathcal A^2 _\lambda$ is a constant that depends on the inclusion geometry and the thermal conductivities of ice and liquid water (\hyperref[sec:theory]{\textcolor{linkcolor}{Methods 6}}). 
This model, analogous to previous derivations \cite{tiller1963migration}, is consistent with experimental observations (Fig.~\ref{fig:figure_NaCl}): migration velocity increases when approaching the ice front, and depends on equilibrium conditions rather than on the amount of salt initially trapped in the inclusion.\\

More precisely, Eq.\eqref{eq:migration_speed} predicts that the migration velocity $v_b$ increases linearly with the temperature gradient $G_i$, and the inverse local temperature $(T_0 - T)^{-1}$. To verify this, we examine the normalised migration velocity $v_b / G_i$ as a function of local temperature in log-log scales (Fig.\ref{fig:figure_NaCl}B). Remarkably, we find that all experiments conducted on NaCl collapse on a single master curve, independently of the imposed temperature gradient and initial salt concentration.\\

Close to the ice front, below a temperature $T_0 - T \sim 1.5\unit{\kelvin}$, the migration velocity deviates from the analytical prediction Eq.~\eqref{eq:migration_speed}, identically for all experimental conditions (Fig.~\ref{fig:figure_NaCl}B). Although we anticipate the origin of this deviation to emanate from a salt accumulation at the rear of the inclusion when diffusive transport cannot keep up with the migration speed, or from the non-planarity of the ice front which would curve the isothermal lines, we leave its precise understanding for future investigation.
\\

In most of the temperature range, however, the migration velocity increases following the $(T_0 - T)^{-1}$ prediction Eq.~\eqref{eq:migration_speed} for all datasets. This result shows that, except for a narrow region close the ice front, brine migration in ice is primarily governed by salt diffusion from the rear end to the front, under a concentration gradient that is slaved to the temperature gradient. We can thus exploit the migration velocity of brine inclusions in ice to estimate the diffusion coefficient of salts in water at negative temperatures. Fitting Eq.~\eqref{eq:migration_speed} for $T_0 - T > 1.5\unit{\kelvin}$ we estimate an effective diffusion coefficient for NaCl of $D = 241\pm8\unit{\square\micro\m}/\unit{\s}$ (\hyperref[sec:fits]{\textcolor{linkcolor}{Methods 7.1}}). 
Although this value is six times lower than previous estimates at room temperature \cite{rard1979}, diffusivity is temperature- and concentration- dependent and our estimates is compatible with extrapolation of literature data to negative temperatures (see \hyperref[sec:fits]{\textcolor{linkcolor}{Methods 7.2}} for a detailed comparison).

\begin{figure}[htbp]
    \centering
    \includegraphics[scale=1]{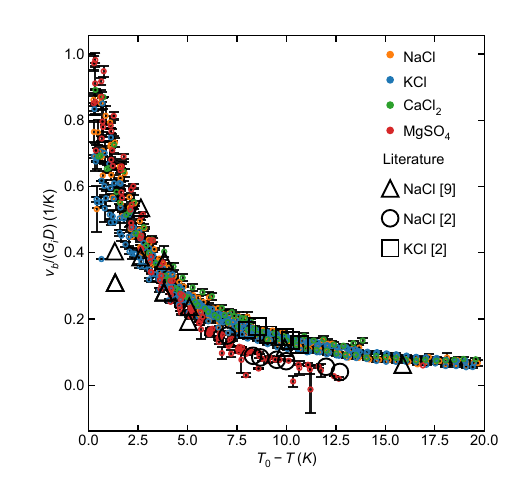}
    \caption{\textbf{Universal scaling of the brine migration dynamics}. Mean normalised velocity $v_b/(G_i D)$ ($1/K$) as a function of $T_0-T$ (K) for NaCl, KCl, CaCl$_2$ and MgSO$_4$. 
    $D$ is the diffusivity fitted independently for each salt from our experimental data (\hyperref[sec:fits]{\textcolor{linkcolor}{Methods 7}}). The empty symbols correspond to the data points from the existing literature~\cite{light2009hydrohalite, hoekstra1965migration}}.
    \label{fig:collapse}
\end{figure}

We applied the same experimental procedure and analysis to three other salts commonly found in seawater, KCl, CaCl$_2$,and MgSO$_4$, and measured their respective diffusion coefficient (\hyperref[sec:fits]{\textcolor{linkcolor}{Methods 7.1}}).
Following the uni-dimensional diffusion model derived above, Eq.~\eqref{eq:migration_speed}, we gather the normalised migration velocity $v_b / (DG_i)$ against local temperature for the four investigated salts and the existing data points from the literature (Fig.~\ref{fig:collapse}).
All data, past and present, remarkably fall onto two branches.

The main branch gathers the vast majority of all data points: present migration velocities of NaCl, KCl, and CaCl$_2$, and past measures for NaCl \cite{light2009hydrohalite} and KCl \cite{hoekstra1965migration}.
It corresponds to an algebraic decay of migration speed with local temperature $v_b / (DG_i) \propto (T_0 - T)^{-1}$, in perfect agreement with migration dynamics controlled by a uni-directional diffusive flux of salts, Eq.~\eqref{eq:migration_speed}. 
The secondary branch gathers present migration velocities of MgSO$_4$ and past measures for NaCl \cite{hoekstra1965migration}, and exhibits a faster decay with decreasing temperature.
Although investigating this deviating behavior is left for future work, we note that the secondary branch coincides with experiments in which migrating inclusions are spherical, a configuration which could break the uni-dimensionality of solute fluxes and lead to different dynamics.

The remarkable collapse of data corresponding to the migration of brine inclusions of different salts, and under temperature gradients spanning three orders of magnitude \cite{light2009hydrohalite, hoekstra1965migration}, is a direct experimental validation of the universal character of brine migration in ice, showing the applicability of our approach to inclusions formed during freezing of real seawater.  
The same diffusion-limited scaling captures the migration dynamics of many different solutes in ice, regardless of their phase diagrams and transport properties.\\

In conclusion, we investigated the migration of hundreds of brine inclusions in ice, under a large range of temperature gradients and initial salt concentration range, for 4 different salts.
We showed that nearly all our measurements and literature data (Fig.~\ref{fig:collapse}) collapse on a single master curve, in agreement with a migration process controlled by salt diffusion inside the inclusions.
This study significantly improves our understanding of brine dynamics, allowing for predictions of brine migration in sea ice knowing the temperature gradients over the year.
Remarkably, the presented experimental setup is a promising method to investigate transport dynamics in solutions at low temperatures.
We demonstrated that the migration velocity informed directly on the diffusion coefficient. Deviations from the evidenced master curve could then provide fundamental information on the migration dynamics of more complex solutes, from polymers to surfactants, or biological condensates.

\bibliographystyle{apsrev4-2}
\bibliography{bibliography} 

\section*{Methods}

\subsection*{1. Experimental setup}
\label{sec:exp_setup}

\textbf{1.1 Sample preparation -- }
Hele-Shaw cells were made from two glass slides separated by a \qty{100}{\micro\meter} gap defined by double-sided adhesive tape. To prepare the saline solutions, we first prepared a stock solution of water and sulforhodamine B (FluoTechnik) at \qty{e-5}{M}. 
We then weighted the required  mass of salt with a precision scale, and added stock solution to reach the targeted salt concentration, either \qty{e-1}, \qty{e-2} or \qty{e-3}{M}. This protocol has the advantage to keep the fluorophore concentration constant in all samples.   The solution was then injected into the cell by capillarity.
\\

\begin{figure}[htbp]
    \centering
    \includegraphics{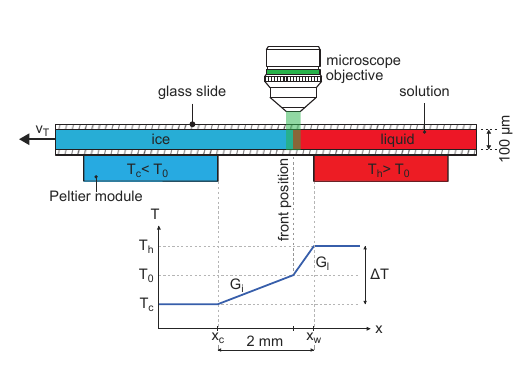}
    \caption{Schematic of the experimental setup (adapted from \cite{dedovets2018temperature}). The saline solution is confined between two glass slides, spaced by 100 \unit{\micro\meter}. We place the sample onto two Peltier modules, at temperatures below and above $T_0$, spaced by a \qty{2}{\milli\meter} gap. This configuration leads to a bi-linear temperature profile in the sample (\hyperref[sec:temp]{\textcolor{linkcolor}{Methods 2.1}}). The sample is attached to a motorized translation stage, used to destablise the ice front and trap brine inclusions in ice. 
    We image the migration dynamics with an upright confocal microscope.
    }
    \label{fig:setup}
\end{figure}

\textbf{1.2 Ice front destabilization -- }
The Hele-Shaw cell was placed onto two fixed Peltier modules (Adaptive, ref. ETC-071-14-11-E), spaced by a \qty{2}{\milli\meter} gap, and attached to a motorized translation stage (Fig.~\ref{fig:setup}).

To generate an ice front inside the cell, we first apply temperatures of \qty{15}{\degreeCelsius} and \qty{-15}{\degreeCelsius} to the Peltier modules on the two sides of the cell.
Once the ice front is stable, we translate the cell at a constant velocity $v_T = 30\unit{\micro\meter}/\unit{s}$ or $50\unit{\micro\meter}/\unit{s}$, depending on the sample, in the direction parallel to the temperature gradient. This process induces a directional freezing that destabilizes the ice front and traps liquid pockets, thereby creating brine inclusions in ice. We pull the sample over \qty{4}{\mm}, which typically traps hundreds of brine inclusions in the ice.
We then change the Peltier temperatures to impose the desired temperature gradient $G_0 =$ \qty{15}, \qty{25}, or $35\unit{\kelvin}/\unit{\milli\meter}$. In this configuration, the sample is subjected to a fixed temperature gradient.
We observe the brine inclusions migrating within the ice towards the warmer temperatures after stabilization of the ice front.
\\

\textbf{1.3 Imaging -- } 
We performed the imaging with a scanning confocal microscope (Leica TCS SP8), exciting the fluorescent dye with a \qty{558}{\nano\meter} laser and collecting the emitted light between 570 and \qty{800}{\nano\meter}. Using a 10x objective, each image covers a  
$\qty{912}{\micro\meter} \times \qty{912}{\micro\meter}$ field of view, with an image size of $512\times512$ pixels, corresponding to a \qty{1.78}{\micro\meter}/pixel resolution. Thanks to this large field of view, we could observe the migration dynamics over a large temperature range. Typical acquisition rate was 0.7 to 0.8 frame per second which was sufficient for effective tracking but not excessive, to avoid collecting overly large dataset. 
\\

\textbf{1.4 Tracking -- }
Detection and tracking is done using the Trackpy \cite{trackpy} python library. 
\subsection*{2. Temperature profiles}
\label{sec:temp}

\textbf{2.1 Temperature profile inside the ice -- }

To estimate the local temperature in the frozen part of the sample, we need to calculate the temperature profile within the ice.
We thus model the experimental setup (Fig.~\ref{fig:setup}) as a one-dimensional system in the direction orthogonal to the ice-liquid interface.
We assume the interface is located at position $x=0$.
The two Peltiers are spaced by a distance $L$, the cold module on the left at $x_c<0$ being fixed at a temperature $T_0 - \Delta T/2$ and the warm module on the right at $x_w = x_c + L > 0$, with a temperature $T_0 + \Delta T/2$. 
There, the freezing temperature of water $T_0 = 273.15\unit{\kelvin}$ is the temperature at the ice front.
Between the two Peltiers the temperature follows the heat equation $\partial_t T - D_{i/l} \partial_{xx}T = 0$, with $D_i$ and $D_l$ the thermal diffusivities of ice and liquid water, respectively.
At the interface, the heat flux continuity gives 
$\lambda_iG_i=\lambda_lG_l$, where $G_i$ and $G_l$ are the temperature gradient in the ice and liquid parts.
\\

In the steady state, the heat equation leads to a piece-wise linear profile $T(x) = T_0 + G_i x$ within the ice, for $x < 0$, $T(x) = T_0 + G_l x$ within the liquid, for  $x > 0$.
The temperature at the ice front being $T_0$, we can express the two gradients as functions of the position $x_c$ of the cold Peltier: 
$G_i = -\Delta T / 2 x_c$ and $G_l = \Delta T / 2 (L + x_c)$.
The conservation of heat flux then leads to $x_c = -L \lambda_i/(\lambda_i + \lambda_l)$ and a temperature gradient inside the ice of
\begin{equation}
    G_i  = \frac{\lambda_i + \lambda_l}{2\lambda_i}G_0,
\end{equation}
where $G_0 = \Delta T / L$ is the imposed temperature gradient.
With a gap length $L = \qty{2}{\milli\meter}$ and thermal conductivities $\lambda_i = \qty{2.2}{\watt}/\unit{\meter}/\unit{\kelvin}$ \cite{slack1980thermal} and $\lambda_l = \qty{0.6}{\watt}/\unit{\meter}/\unit{\kelvin}$ \cite{Takizawa1978,Huber2012} we obtain $x_c = \qty{-1.57}{\milli\meter}$ and a temperature gradient $G_i = 0.63\,G_0$.
\\

\noindent
Experimentally, we obtain the local temperature from the inclusion position $x$ following $T_0 - T(x) = - G_i x$.\\

\textbf{2.2 Temperature profile inside the inclusion --}
As a brine inclusion perturbs the local temperature field, the temperature gradient within it $G_b$ differs from the temperature gradient within the ice $G_i$. 
To account for this change, we use the solution outlined by Tiller \cite{tiller1963migration}

\begin{equation}
    G_b = \mathcal A_\lambda G_i - \mathcal B_\lambda\, \rho_i\mathcal L\, v_b
    \label{eq:GradientInclusion}
\end{equation}
where $v_b$ is the migration velocity in the bin, $\rho_i=  920\,\unit{\kilo\gram}/\unit{\meter\cubed}$ the density of ice \cite{LAAKSONEN202231}, and $\mathcal L \approx 3.10^{5}\,\unit{\joule}/\unit{\kilo\gram}$ the mass latent heat of fusion. The first term accounts for the difference in thermal conductivities, providing a similar correction as the one described above.
The second term accounts for the exchange of latent heat at the moving interface. 
Both terms vary depending on the inclusion geometry through the constants $\mathcal A_\lambda$ and $\mathcal B_\lambda$.
For instance, $\mathcal A_\lambda = 3\lambda_i / (2\lambda_i + \lambda_l)$ and $\mathcal B_\lambda = 1 / (2\lambda_i + \lambda_l)$ for a spherical inclusion \cite{you2021thermal, tiller1963migration}, while $\mathcal A_\lambda = 2\lambda_i / (\lambda_i + \lambda_l)$ and $\mathcal B_\lambda = 1 / (\lambda_i + \lambda_l)$ for a cylindrical inclusion \cite{tiller1963migration}.

In our experimental setup, the ratio of both terms $\mathcal A_\lambda G_i /  \mathcal B_\lambda\, \rho_i\mathcal L\, v_b > 10^2$ is such that we neglect the second term and further consider %

\begin{equation}
    G_b = \mathcal A_\lambda\frac{(\lambda_i + \lambda_l)}{2\lambda_i}G_0. 
    \label{eq:GradientInclusion_Simplifiee}
\end{equation}

\subsection*{3. Velocities binning}
\label{sec:vel}

\begin{figure}[htbp]
    \centering
    \includegraphics{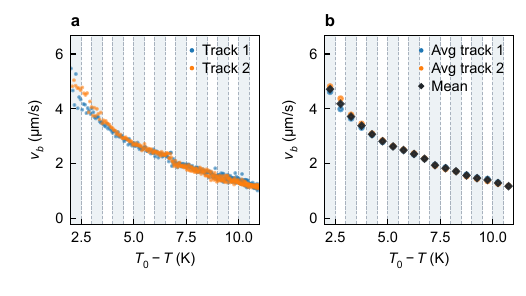}
    \caption{Schematic of the averaging method. (a) Migration velocity as a function of local temperature for two individual brine inclusions. (b) Averaged migration velocity as a function of local temperature. Orange and blue circles: averaged migration velocity within each temperature bin for individual trajectories. Black diamonds: velocities averaged across trajectories with equal weight. Shaded bands indicate temperature bins.}
    \label{fig:avg_example}
\end{figure}

Instantaneous inclusion velocities were calculated from frame-to-frame displacements of the tracked inclusions. 
To obtain the averaged migration velocities while avoiding over-weighting long trajectories, we proceeded as following.
First, we sliced the temperature axis in bins of $0.5 \unit\kelvin$ (Fig.\,\ref{fig:avg_example}a).
Second, for each individual trajectory, we computed the velocity averaged within each bin (Fig.\,\ref{fig:avg_example}b, colored circles).
At last, we averaged these velocities over all trajectories in each bin (Fig.\,\ref{fig:avg_example}b, black diamonds). 

The standard error of the mean was calculated from the distribution of the trajectory-averaged velocities in each bin. 

\subsection*{4. Phase change dynamics}
\label{sec:phasechange}

Along the inclusion boundaries, the melting and thawing speed $v_{\varphi}$ is controlled by the difference between the local concentration and the equilibrium concentration, $v_{\varphi} = \mu (c(T) - c^*(T)) / \alpha$.
The liquidus slope $\alpha$ is of the order of $\qty{0.2}{M}/\unit{\kelvin}$ \cite{koop2000phase} and the atomic kinetic coefficient $\mu$ of $10^{-2}$ to $\qty{10}{\meter}/\unit{\second}/\unit{\kelvin}$ \cite{tiller1963migration}.
We estimate the characteristic time associated with phase change by considering a small displacement $\delta r$ of one liquid-ice boundary away from its local equilibrium position. This displacement changes the local equilibrium concentration by $\delta c \sim \alpha G_b \delta r$. The corresponding phase-change velocity is therefore $v_{\varphi} \sim \mu G_b \delta r$. We can then define the typical relaxation time of such a perturbation as 

\begin{equation}
    \tau_{\varphi} = \frac{\delta r}{v_{\varphi}} = \frac{1}{\mu G_b}
\end{equation}
and we observe that it does not depend on the local temperature nor the concentration in the linear approximation. This time has to be compared with the characteristic migration time of a brine inclusion over its own size $L_b$
\begin{equation}
    \tau_{mig} \sim \frac{L_b}{v_b}
\end{equation}

Considering the value given by Tiller~\cite{tiller1963migration} and $G_b \sim 10^4 \unit{\kelvin}/\unit{\meter}$, we get $\tau_{\varphi ,min} \sim 10^{-5} \unit{\second}$ and $\tau_{\varphi, max} \sim 10^{-2} \unit{\second}$ to compare with, using the value of size and migration velocity mentioned in the paper, $\tau_{mig, min} \sim 1 \unit{\second}$. 

In the most unfavorable case, we have 
\begin{equation}
    \frac{\tau_{\varphi,max}}{\tau_{mig,min}} \sim 10^{-2} 
\end{equation}
telling that the typical time of melting/freezing is in the worst case 100 times faster the time it takes to a brine inclusion to migrate over its size.

\subsection*{5. Radial diffusive time}
\label{sec:radial}

The transverse diffusion time is $\tau_D^{\perp} \sim \ell_b^2/D$ with $\ell_b$ the transverse size of the inclusion. We thus define a transverse Peclet number
 
\begin{equation}
    Pe_{\perp} \sim \frac{\tau_D^{\perp}}{\tau_{mig}} \sim \frac{v_b L_b}{D} \left( \frac{\ell_b}{L_b} \right)^2
\end{equation}

Since $v_bL_b/D$ is at most of order of unity and in our inclusions $\ell_b/L_b < 1$, then $Pe_{\perp} \ll 1$ so that transverse diffusion is fast compared with the migration dynamics. Any salt concentration heterogeneity that may develop across the brine inclusion is therefore rapidly smoothed out before the inclusion migrates over its own length, supporting a one-dimensional description along the migration direction.

\subsection*{6. Conservation law and migration velocity}
\label{sec:theory}

As mentioned in the main text, we assume the concentration profile is slaved to the linear temperature profile $c = c^*(T)$ with $T(x, x_b) = T_b + G_b(x - x_b)$.
Noting that the temperature at the center of the inclusion is $T_b = T_0 + G_i x_b$, and that the inclusion moves in time following $\dot x_b = v_b$, 
the conservation law writes
\begin{equation}
    \left((G_i - G_b)v_b + G_b v_x\right)\frac{dc^*}{dT} - D G_b^2 \frac{d^2c^*}{dT^2} = 0\nonumber,
\end{equation}
where we assumed that the diffusion coefficient was a constant and we used the incompressibility in the absence of transverse flow $\partial_x v_x = 0$.
Assuming that the equilibrium concentration varies linearly with temperature $c^*(T) = \alpha(T_0 - T)$, the advection velocity writes $v_x= v_b (G_b - G_i) / G_b$.\\

Further assuming the inclusion length does not vary in time, a balance between the salt flux inside the inclusion $j_c  = v_x c - D\partial_x c$ and the salt transported due to the migration $c^*(T)v_b$ leads to
\begin{equation}
    (T_0 - T)v_b \left(\frac{G_b - G_i}{G_b}\right) + D G_b = (T_0 - T) v_b\nonumber,
\end{equation}
and to
\begin{equation}
    v_b = \frac{G_b^2}{G_i}\frac{D}{T_0 - T}\nonumber.
\end{equation}
Leveraging the analysis developed on temperature profiles in (\hyperref[sec:temp]{\textcolor{linkcolor}{Methods 2}}) we obtain
\begin{equation}
    \frac{v_b}{G_i} = \mathcal A^2_{\lambda}\frac{D}{T_0 - T}\nonumber.
\end{equation}
The prefactor 
$\mathcal A_\lambda^2$ 
is of order 1 and depends on the inclusion geometry (\hyperref[sec:temp]{\textcolor{linkcolor}{Methods 2.2}}).

\subsection*{7. Diffusion coefficient}
\label{sec:fits}

\textbf{7.1 Estimation of the diffusion coefficient}

To estimate a diffusion coefficient value for each salt, we assumed the migration velocities followed an algebraic decay with local temperature Eq.\eqref{eq:migration_speed}.
More specifically, for each individual experiment we converted the binned velocity and local temperature in logarithmic scales, and fitted them with a linear law of fixed slope $\log (v_b/G_i) = C - \log(T_0 - T)$  over the range $T_0 - T > 1.5\unit{\kelvin}$.

\begin{figure}[htbp]
    \centering
    \includegraphics[width=89mm]{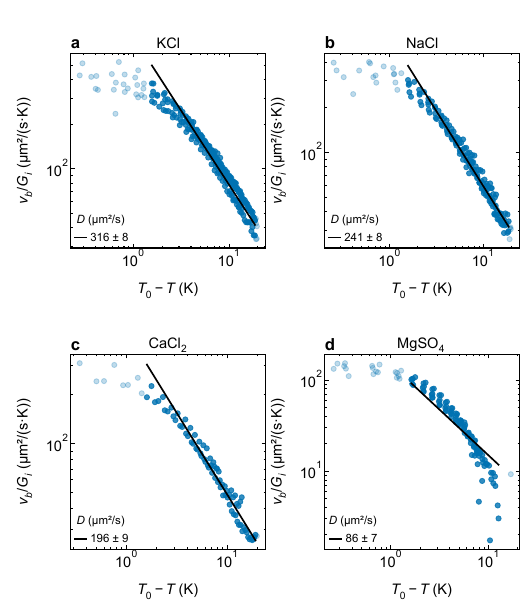}
    \caption{Normalised migration velocity $v_b / G_i$ against local temperature for (a) KCl, (b) NaCl, (c) CaCl$_2$, and (d) MgSO$_4$. Blue circles correspond to the binned data of all individual experiments. Black lines correspond to an algebraic decay $\log(v_g / G_i) = \log(\mathcal A^2_\lambda D) - \log(T_0 - T)$ with the diffusivities reported in Table~\ref{tab:diffusivity_table}.
    }
    \label{fig:diffusivity}
\end{figure}

We extracted the diffusivity from the fitted coefficient $C = \log(\mathcal A_\lambda^2D)$, assuming a spherical inclusion formula for MgSO$_4$ and a cylindrical one for KCl, NaCl, and CaCl$_2$ (\hyperref[sec:temp]{\textcolor{linkcolor}{Methods 2.2}}).
We then averaged the diffusivities over all experiments for each salt.
We report the averaged values graphically in Fig.~\ref{fig:diffusivity}, and numerically in Table~\ref{tab:diffusivity_table}.
As mentionned in the main text, we note that the migration velocity for the MgSO$_4$ inclusions deviate from a $(T_0 - T)^{-1}$ decay.\\

\begin{table}[htbp]
    \centering
    \small
    \setlength{\tabcolsep}{5pt}
    \renewcommand{\arraystretch}{1.2}

    \makebox[\linewidth][c]{%
    \begin{tabular}{cccccc}
        \hline
        \makecell[cc]{Salt}
        & \makecell[cc]{$D_{\mathrm{eff}}$ ($\unit{\micro\m\squared}/\unit{s}$)}
        & \makecell[cc]{$D_{\mathrm{extr}}^{253\unit{\K}}$ ($\unit{\micro\m\squared}/\unit{s}$)}
        & \makecell[cc]{$D_{\mathrm{extr}}^{263\unit{\K}}$ ($\unit{\micro\m\squared}/\unit{s}$)}
        & \makecell[cc]{$D_{\mathrm{extr}}^{273\unit{\K}}$ ($\unit{\micro\m\squared}/\unit{s}$)}
        & \makecell[cc]{Reference}\\
        \hline
        KCl       & $316 \pm 8$ & $582$ & $788$ & $1043$ & \cite{castillo1993}\\
        NaCl      & $241 \pm 8$ & $436\vert435$ & $607\vert588$ & $824\vert778$ &  \cite{castillo1993}$\vert$\cite{caldwell1981soret}\\
        CaCl$_2$  & $196 \pm 9$ & $85$ &  $163$ & $298$ &  \cite{caldwell1981soret}\\ 
        MgSO$_4$  & $86 \pm 7$ & $147$ &  $209$ & $289$ &  \cite{caldwell1981soret}\\
        \hline
    \end{tabular}%
    }
    \caption{Effective diffusivities obtained from the migration velocity of brine inclusions, and extrapolated diffusion coefficients $253$, $263$, and $273\unit{\K}$ for the four salts we investigated. The effective diffusivity correspond to the average over all experiments for a single salt, and the uncertainty to the standard error of the mean.}
    \label{tab:diffusivity_table}
\end{table}

\textbf{7.2 Extrapolation of literature data to negative temperatures}

\begin{figure}[htbp]
    \centering
    \includegraphics[width=89mm]{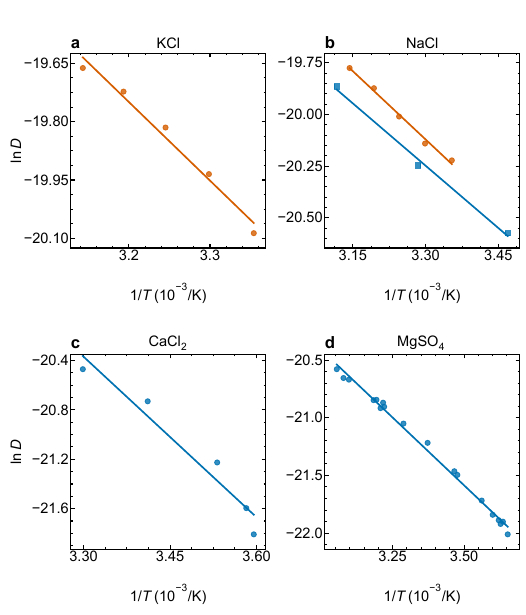}
    \caption{Diffusion coefficient against inverse temperature, in log-lin scale, for (a) KCl, (b) NaCl, (c) CaCl$_2$, and (d) MgSO$_4$. The orange and blue circles respectively correspond to literature data from Castillo \& Garza \cite{castillo1993}, and Caldwell \& Eide \cite{caldwell1981soret}.
    The full lines correspond to the fits we obtained with an Arrhenius law Eq.~\eqref{eq:arrhenius}.
    }
    \label{fig:extrapolation}
\end{figure}

To the best of our knowledge, no literature data reports salt diffusion coefficients at temperatures below the freezing point.
To compare our estimate with literature data, we thus extrapolated literature measurements at different temperatures above the freezing point for KCl\cite{castillo1993}, NaCl\cite{castillo1993,caldwell1981soret}, CaCl$_2$\cite{caldwell1981soret}, and MgSO$_4$\cite{caldwell1981soret} with an Arrhenius law \cite{korber1988extrapolation}

\begin{equation}
    D(T)=D_a\exp\left(-\frac{E_a}{k_BT}\right),
    \label{eq:arrhenius}
\end{equation}
where $T$ is the absolute temperature, $D_a$ a prefactor, $E_a$ the activation energy, and $k_B$ the Boltzmann constant.
We obtained the fitting parameters ($D_a, E_a$) from a linear regression of $\ln D(T)$ against $1/T$ (Fig.~\ref{fig:extrapolation}), and we report the extrapolated diffusion coefficients at $253$, $263$, and $273\unit{\K}$ for four salts in Table~\ref{tab:diffusivity_table}.
Although this method does not account for concentration- or pressure- dependence of the diffusion coefficient, we note that the values estimated from the migration of brine inclusions are close to the extrapolated ones.

\section*{Supplementary Movie}

Migration of KCl brine inclusions through ice. The KCl solution initial concentration is $c_0 = \qty{e-2}{M}$, and the imposed temperature gradient is $G_0 = 15\unit{K}/\unit{mm}$. The time stamp is expressed in minutes and seconds.
\clearpage

\section*{Data and code availability}
Data and code will be made available in a public GitHub repository upon publication. We can of course provide them to the editors/reviewers if needed.

\section*{Funding} 
We acknowledge the Materials Science Doctoral School (EDML) of the Université Lyon 1 for the financial support through a PhD fellowship and the ANR - FRANCE (French National Research Agency) for its financial support of the FROST project ANR-22-CE06-0022.

\begin{acknowledgments}
We thank Gilles Simon for his help on the cryo-confocal, as well as Christophe Ybert, Francesco Picella, and Thomas Guérin for useful discussions.
\end{acknowledgments}

\section*{CRediT author statement: }
\textbf{Natanaël Braos--d'Urso}: Investigation, Validation, Formal analysis, Writing - Original Draft, Visualization
\textbf{Nicolas Bain}: Conceptualization, Writing - Review \& Editing, Supervision
\textbf{Bruno Issenmann}: Conceptualization, Writing - Review \& Editing, Supervision
\textbf{Sylvain Deville}: Conceptualization, Writing - Review \& Editing, Supervision

\section*{Conflict of interest} 
\label{sec:conflict_of_interest}
The authors declare no conflicts of interest.

\section*{Correspondence}
Correspondence and requests for materials should be addressed to Sylvain Deville (sylvain.deville@univ-lyon1.fr)

\end{document}